# Study of quality assurance regulations for linear accelerators in Korea: A comparison study between the current status in Korea and the international guidelines

Hyunho Lee, Seonghoon Jeong, Yunhui Jo and Myonggeun Yoon*

*Department of Bio-convergence Engineering, Korea University, Seoul, Korea.*

Quality assurance (QA) for medical linear accelerators is indispensable for appropriate cancer treatment. Some international organizations and western advanced countries provide QA guidelines for linear accelerators. Currently, QA regulations for linear accelerators in Korean hospitals specify a system in which each hospital stipulates its independent hospital-based protocols for QA procedures (HP_QAPs) and conducts QA based on these HP_QAPs while regulatory authorities verify whether items under these HP_QAPs have been performed. However, because this regulatory method cannot guarantee the quality of universal treatment, and QA items with tolerance criteria are different in many hospitals, the presentation of standardized QA items and tolerance criteria is essential. In this study, QA items in HP_QAPs from various hospitals and those presented by international organizations, such as the International Atomic Energy Agency, European Union, and American Association of Physicist in Medicine, and western advanced countries, such as the USA, the UK, and Canada, were compared. Concordance rates between QA items for linear accelerators that were presented by the aforementioned organizations and those currently being implemented in Korean hospitals were shown to exhibit a daily QA of 50%, a weekly QA of 22%, a monthly QA of 43%, and an annual QA of 65%, and the overall concordance rates of all QA items were approximately 48%. In comparison between QA items being implemented in Korean hospitals and those being implemented in western advanced countries, concordance rates were shown to exhibit a daily QA of 50%, a weekly QA of 33%, a monthly QA of 60%, and an annual QA of 67%, and the overall concordance rate of all QA items were approximately 57%. The results of this study indicate that the HP_QAPs currently implemented by Korean hospitals as QA standards for linear accelerators used in radiation therapy do not meet international standards. To solve this problem, it is necessary to develop national standardized QA items and procedures for linear accelerators.

Email: radioyoon@korea.ac.kr

Fax: +82-2-940-2823

# I. INTRODUCTION

Cancer treatment aims to provide sufficient radiation doses for tumors with minimum doses to normal tissues, thereby completely curing cancers or improving the quality of life of patients by reducing the volumes of, or suppressing the growth of, lesion tissues. To this end, advanced radiotherapy techniques such as intensity-modulated radiation therapy, image-guided radiation therapy, and stereotactic radiotherapy have been introduced into radiotherapy using a linear accelerator, which is a radiotherapy device, and these advanced radiotherapy techniques require high degrees of precision of radiotherapy machines. As a result, these machines should be periodically checked against electric malfunction and mechanical failure through quality assurance (QA) for linear accelerators.

The International Atomic Energy Agency (IAEA) presents the items, trial cycle, and tolerance of QA for linear accelerators [1]. QA procedures are also presented in the European Commission Radiation Protection No.91 of the EU and the task group 40 (TG-40) and TG-142 reports of the American Association of Physicists in Medicine (AAPM) [2-4]. Not only international organizations but also western advanced countries such as the USA and Canada in North America and the UK and Switzerland in Europe present standardized items, trial cycles, and tolerance levels of QA for linear accelerators through reports and papers, and QA teams of individual hospitals autonomously determine QA procedures based on the foregoing discussion [5-10].

Currently, each hospital in Korea submits hospital-based protocols for QA procedures (HP_QAPs) to the Korea Institute of Nuclear Safety (KINS) to be approved for the treatment of patients using linear accelerators, and audits are later conducted to evaluate how the determined items are implemented on the basis of these HP_QAP items. The items, trial cycles, and tolerance levels of HP_QAPs are autonomously set by each hospital based on its QA teams. While international organizations and western advanced countries present the guideline of QA procedures for linear accelerators, Korea has reached no consensus about standardized QA indicators for linear accelerators. Furthermore, the present state of the HP_QAPs of hospitals in Korea has been neither investigated nor analyzed in comparison with the present state of QA for linear accelerators overseas.

This study aims to investigate the current status of QA regulations for linear accelerators to be conducted through HP_QAPs of 18 hospitals in Korea and analyze common features and differences of the HP_QAPs in comparison with QA items for linear accelerators presented by international organizations and western advanced countries to obtain data for the development of standardized QA indicators for linear accelerators.

## II. METHODS AND MATERIALS

The present state of HP_QAPs in Korea was investigated by requesting and collecting HP_QAP data submitted by 18 Korean hospitals to the KINS before the introduction of linear accelerator for the treatment of patients. The data obtained were then compared with the present state of QA procedures for linear accelerators presented by three international organizations—the IAEA, EU, and AAPM—and four western advanced countries—the USA, Canada, the UK, and Switzerland.

The present state of HP_QAPs being implemented by the 18 Korean hospitals is listed in Tables 1–4. The HP_QAPs were divided into daily QA, weekly QA, monthly QA, and annual QA according to QA trial cycles for linear accelerators, and the implementation rates of individual items were set to indicate the ratios of hospitals that were implementing QA for linear accelerators for the relevant items to the 18 hospitals (For example, if six hospitals out of the 18 hospitals conduct inspections for item 1, the implementation rate of item 1 is 33%.).

As for the QA procedures for linear accelerators presented by international organizations, handbooks and reports issued by the IAEA, EU, and AAPM were referred to in order to set table items, and the procedures were divided into daily QA, weekly QA, monthly QA, and annual QA according to QA trial cycles for linear accelerators. To set QA procedures for linear accelerators presented by international organizations, items being implemented by at least two of the three organizations—the IAEA, EU, and AAPM—were collected, and a table was constructed.

The QA procedures for linear accelerators presented by western advanced countries were set on the basis of those implemented in the USA and Canada in North America and the UK and Switzerland in Europe. The QA procedures for linear accelerators were divided into daily QA, weekly QA, monthly QA, and annual QA according to QA trial cycles for linear accelerators, and a table of QA procedures for linear accelerators presented by western advanced countries based on QA items presented by at least two of the four countries was constructed.

To compare the present state of QA procedures for linear accelerators followed overseas with that of HP_QAPs in Korea, hospital-independent QA regulations for linear accelerators in Korea *vs.* QA procedures for linear accelerators by international organizations were compared, as well as hospital-independent QA regulations for linear accelerators in Korea *vs.* QA procedure for linear accelerators in western advanced countries. The results were indicated as items, tolerance levels, and implementation rates, which are the ratios of items being commonly implemented in the HP_QAPs in Korean hospitals to QA procedure items for linear accelerators followed overseas.

## III. RESULTS AND DISCUSSION

Tables 5–8 show the results of the comparison between the present state of QA procedures for linear accelerators presented by international organizations with that of HP_QAPs in Korea. The "0%" in execution rate means that the corresponding item is not included in the HP_QAPs of surveyed18 hospitals in Korea. Among eight daily QA items for linear accelerators presented by international organizations, five were being implemented as HP_QAP items in at least one hospital in Korea, while four were being implemented as HP_QAP items in the majority of hospitals in Korea, which showed an implementation rate exceeding 50%. In the case of weekly QA, among nine items presented by international

organizations, four were being implemented as HP_QAP items in at least one hospital in Korea, while two were being implemented as HP_QAP items in the majority of hospitals in Korea, which showed a 22% implementation rate. A total of 22 monthly QA items for linear accelerators were recommended by international organizations; sixteen of them were being implemented as HP_QAP items in at least one hospital in Korea, while four were being implemented as HP_QAP items in the majority of hospitals in Korea, which showed a 43% implementation rate. Finally, in the case of annual QA, among 20 items presented by international organizations, 17 were being implemented as HP_QAP items in at least one hospital in Korea, while 13 were being implemented as HP_QAP items in the majority of hospitals in Korea, showing an implementation rate exceeding 65%.

Tables 9–12 show the results of comparison between the present state of QA procedures for linear accelerators presented by western advanced countries and that of HP_QAPs in Korea. A total of 10 daily QA items for linear accelerators were presented by western advanced countries; six of them were being implemented as HP_QAP items in at least one hospital in Korea, while five were being implemented as HP_QAP items in the majority of hospitals in Korea, which showed an implementation rate exceeding 50%. In the case of weekly QA, among six items presented by western advanced countries, two were being implemented as HP_QAP items in at least one hospital in Korea, and two were also being implemented as HP_QAP items in the majority of hospitals in Korea, which showed an implementation rate of 33%. Among 20 monthly QA items for linear accelerators presented by western advanced countries, 16 were being implemented as HP_QAP items in at least one hospital in Korea, while 12 were being implemented as HP_QAP items in the majority of hospitals in Korea, which showed an implementation rate of 60%. In the case of annual QA, among 18 items presented by western advanced countries, 16 were being implemented as HP_QAP items in at least one hospital in Korea, while 12 were being implemented as HP_QAP items in the majority of hospitals in Korea, which showed an implementation rate exceeding 67%.

To compare all QA procedure items for linear accelerators presented by international organizations and those being implemented in Korea, a total of 58 QA items for linear accelerators were presented by international organizations; of these items, 28 were being implemented as HP_QAP items in the majority of hospitals in Korea, which showed a 48% implementation rate. To compare all the HP_QAP items presented by western advanced countries and those being implemented in Korea, a total of 54 QA items for linear accelerators were presented by western advanced countries; of these items, 31 were being implemented as HP_QAP items in the majority of hospitals in Korea, which showed a 57% implementation rate. On comparison of HP_QAPs being implemented by hospitals in Korea with QA procedures for linear accelerators presented by international organizations and western advanced countries by the foregoing results, HP_QAPs being implemented by the majority of hospitals in Korea showed an approximately 50% implementation rate, indicating that HP_QAPs implementation in Korea should be commended.

In addition, on reviewing the results of the comparison between HP_QAPs being implemented by hospitals in Korea and QA procedures for linear accelerators presented by international organizations and western advanced countries, compared with those of daily, monthly, and annual QA, the implementation rates of weekly QA were lower with 22% and 33% of the QA procedures for linear accelerators presented by international organizations and western advanced countries, respectively. This can be attributed to the fact that the QA procedures for linear accelerators presented by the AAPM's TG-142 report did not specify weekly QA, and it can be inferred that QA procedures for linear accelerators in hospitals in Korea were intensively established with reference to the TG report by the AAPM [2-3].

On reviewing the numbers of monthly QA items in the HP_QAPs collected from the 18 hospitals, the numbers were diverse by hospital ranging from a minimum of 12 to a maximum of 27. This diversity in HP_QAPs in Korea is attributable to the fact that HP_QAPs used as reference data for checking radiotherapy in hospitals are autonomously set by QA teams of hospitals. If judged only based on the number of HP_QAP items, some hospitals may be judged as not properly implementing QA for linear accelerators. However, HP_QAPs involve regulated items submitted to the KINS for regulation and can be different from the QA for linear accelerators actually implemented by hospitals. QA for linear accelerators is flexibly implemented depending on situations, and more various items of QA for linear accelerators can be implemented than HP_QAP items.

Because of the nature of HP_QAPs for which regulated items are to be autonomously determined, there is always a possibility to implement fewer QA items, as compared with all the QA items, to be set as HP_QAP items for reducing the number of items under regulation. Of course, there is no problem if the number of HP_QAP items is small; however, the actual QA for linear accelerators is implemented for more diverse items. However, if QA for linear accelerators is implemented according to the minimized HP_QAPs, ideal radiotherapy effects cannot be expected, and this is closely related to national health. This problem originates from the lack of national standardized QA procedures and can be solved through the development of such QA guidelines.

## IV. CONCLUSIONS

In this study, before developing standardized QA procedure guidelines for linear accelerators in Korean hospitals, hospital-independent QA regulations for linear accelerators of 18 hospitals in Korea were collected and analyzed to investigate and compare the present state of QA procedures with that recommended by international organizations—the IAEA, EU, and AAPM—and western advanced countries—the USA and Canada in North America and the UK and Switzerland in Europe. The results of this study are considered usable as seed data for the development of standardized Korean QA guidelines for linear accelerators.

## ACKNOWLEDGEMENT

This work was supported by the Nuclear Safety Research Program (Grant No. 1305033) through the Korea Radiation Safety Foundation (KORSAFe) and the Nuclear Safety and Security Commission (NSSC), Republic of Korea and the National Nuclear R&D Program through the National Research Foundation of Korea (NRF), funded by the Ministry of Education, Science and Technology (NRF-2013M2A2A7067089) and by Basic Science Research Program through the National Research Foundation of Korea(NRF) funded by the Ministry of Education, Science and Technology (NRF-2013R1A1A2007630).

Table 1. The daily QA procedure of 18 hospitals in Korea

| Procedure | Execution rate (%) |
|---|---|
| Radiation ON/OFF switches | 94.1 |
| Beam on indicator | 94.1 |
| Door closing safety | 94.1 |
| Audiovisual monitors | 100.0 |
| Emergency off switches | 88.2 |
| Mechanical check (ODI, laser, degree of vacuum, warning light) | 64.7 |
| Environmental check of the treatment room (thermometer, barometer, cleaning) | 47.1 |
| Light/radiation field coincidence | 11.8 |
| Gantry rotational isocenter & gantry angle indicators | 5.9 |
| X-ray flatness and symmetry | 5.9 |
| Electron flatness and symmetry | 5.9 |
| Operational check of MLC | 17.6 |
| Operational check of EPID | 17.6 |
| X-ray / electron output constancy | 47.1 |
| CBCT IC check | 5.9 |

Table 2. The weekly QA procedure of 18 hospitals in Korea

| Procedure | Execution rate (%) |
|---|---|
| Radiation ON/OFF switches | 16.7 |
| Beam on indicator | 16.7 |
| Door closing safety | 16.7 |
| Audiovisual monitors | 16.7 |
| Emergency off switches | 16.7 |
| Gantry angle indicators | 83.3 |
| Collimator angle indicators | 100.0 |
| Couch angle indicators | 100.0 |
| Couch travel in all directions | 16.7 |
| ODI test | 50.0 |
| Accessory Tray | 16.7 |
| Light field size indicator | 66.7 |
| Laser localization | 50.0 |
| Operational check of MLC | 16.7 |
| X-ray output constancy | 83.3 |
| Electron output constancy | 83.3 |

Table 3. The monthly QA procedure of 18 hospitals in Korea

| Procedure | Execution rate (%) |
|---|---|
| Emergency off switches | 61.1 |
| Radiation ON/OFF switches | 27.8 |
| Beam on indicator | 22.2 |
| Door closing safety | 38.9 |
| Audiovisual monitors | 27.8 |
| Collimator rotation isocenter and angle indicator | 94.4 |
| Collimator angle indicator | 5.6 |
| Gantry angle indicator | 5.6 |
| Gantry rotation isocenter | 94.4 |
| Couch travel in all directions | 11.1 |
| Couch rotation isocenter and angle indicator | 94.4 |
| Light field rotation isocenter and size indicator | 33.3 |
| Light/radiation field coincidence | 55.6 |
| Latching of wedges | 16.7 |
| ODI test | 55.6 |
| Laser localization | 50.0 |
| Leaf position accuracy | 16.7 |
| Asymmetric Jaw test | 27.8 |
| picket Fence test | 11.1 |
| X-ray output constancy | 83.3 |
| X-ray flatness and symmetry | 66.7 |
| X-ray dose rate constancy | 16.7 |
| X-ray MU Linearity | 5.6 |
| X-ray energy constancy | 33.3 |
| Electron output constancy | 83.3 |
| Electron flatness and symmetry | 66.7 |
| Electron dose rate constancy | 11.1 |
| Electron MU Linearity | 5.6 |
| Electron energy constancy | 38.9 |
| Operational check of EPID | 11.1 |
| Backup monitor constancy | 5.6 |
| Stability of OBI | 11.1 |
| Accuracy of the mechanical location of OBI and kV source | 11.1 |
| Coincidence of OBI and kV source axis | 11.1 |
| Leakage dose measurement | 33.3 |

Table 4. The annual QA procedure of 18 hospitals in Korea

| Procedure | Execution rate (%) |
|---|---|
| Emergency off switches | 16.7 |
| Radiation ON / OFF switches | 16.7 |
| Beam on indicator | 16.7 |
| Door closing safety | 16.7 |
| Audiovisual monitors | 16.7 |
| Collimator rotation isocenter and angle indicators | 66.7 |
| Collimator radiation isocenter | 50.0 |
| Collimator leakage | 16.7 |
| Collimator Jaw position indicators | 16.7 |
| Gantry rotation isocenter and angle indicator | 50.0 |
| Gantry angle indicators | 16.7 |
| Gantry radiation isocenter | 50.0 |
| Couch angle indicators | 50.0 |
| Couch horizontality and location | 50.0 |
| Table top sag | 33.3 |
| Couch rotation isocenter | 33.3 |
| Couch radiation isocenter | 50.0 |
| Couch motion interlock | 33.3 |
| Couch z-axis movement accuracy | 50.0 |
| ODI test | 33.3 |
| Laser location | 16.7 |
| Wedge factor constancy | 50.0 |
| Monitor chamber linearity | 16.7 |
| Light field size | 33.3 |
| Light field divergence and cross-hair centering | 16.7 |
| Light/radiation field coincidence | 33.3 |
| Asymmetric Jaw position indicators | 50.0 |
| X-ray output constancy | 16.7 |
| X-ray energy constancy | 100.0 |
| X-ray dose correction | 16.7 |
| X-ray short-term and long term output constancy | 16.7 |
| X-ray dose rate output constancy | 50.0 |
| X-ray gantry angle output constancy | 50.0 |
| X-ray flatness and symmetry | 83.3 |
| X-ray FWHM test | 33.3 |
| X-ray output constancy with field size | 50.0 |
| X-ray transmission factor of wedge and accessary | 33.3 |
| X-ray MU linearity | 50.0 |
| Electron output constancy | 16.7 |
| Electron energy constancy | 83.3 |
| Electron dose correction | 16.7 |
| Electron applicator interlock | 16.7 |
| Electron output constancy with applicators | 50.0 |
| Electron flatness and symmetry | 83.3 |
| Electron short-term and long term output constancy | 16.7 |
| Electron dose rate output constancy | 33.3 |
| Electron output constancy with gantry angle | 50.0 |
| Electron FWHM test | 33.3 |
| Electron MU linearity | 50.0 |
| End effect check | 33.3 |
| Independence dose rate test | 16.7 |
| Leakage dose measurement | 16.7 |
| TPS data reconfirm | 16.7 |

Table 5. The comparison between the present state of QA procedures for linear accelerators presented by international organizations with that of HP_QAPs in Korea (daily QA)

| Procedure | Tolerance | Execution rate of the international organization (%) | Execution rate of the hospital in Korea (%) |
|---|---|---|---|
| X-ray output constancy | 3% | 100 | 0 |
| Electron output constancy | 3% | 100 | 0 |
| Localizing lasers | 2 mm | 100 | 64.7 |
| Optical distance indicator (ODI) | 2 mm | 100 | 64.7 |
| Door interlock | Functional | 100 | 94.1 |
| Audiovisual monitor | Functional | 100 | 100 |
| Field size indicator | 2 mm | 66.7 | 11.8 |
| Record | Performance | 66.7 | 0 |

Table 6. The comparison between the present state of QA procedures for linear accelerators presented by international organizations with that of HP_QAPs in Korea (weekly QA)

| Procedure | Tolerance | Execution rate of the international organization (%) | Execution rate of the hospital in Korea (%) |
|---|---|---|---|
| X-ray output constancy | 2% | 100 | 83.3 |
| Electron output constancy | 2% | 100 | 83.3 |
| Backup monitor constancy | 2% | 66.7 | 0 |
| X-ray central axis dosimetry parameter constancy (PDD, TAR, TPR) | 2% | 66.7 | 0 |
| Electron central axis dosimetry parameter constancy (PDD) | 2 mm @ therapeutic depth | 66.7 | 0 |
| X-ray beam flatness constancy | 2% | 100 | 0 |
| Electron beam flatness constancy | 3% | 100 | 0 |
| Wedge factor constancy | 3% | 66.7 | 16.7 |
| Emergency off switches | Functional | 66.7 | 16.7 |

Table 7. The comparison between the present state of QA procedures for linear accelerators presented by international organizations with that of HP_QAPs in Korea (monthly QA)

| Procedure | Tolerance | Execution rate of the international organization(%) | Execution rate of the hospital in Korea (%) |
|---|---|---|---|
| X-ray output constancy | 2% | 100 | 83.3 |
| Electron output constancy | 2% | 100 | 83.3 |
| Backup monitor constancy | 2% | 66.7 | 0 |
| X-ray central axis dosimetry parameter constancy (PDD, TAR, TPR) | 2% | 100 | 33.3 |
| Electron central axis dosimetry parameter constancy (PDD) | 2 mm @ therapeutic depth | 100 | 33.3 |
| X-ray beam flatness constancy | 2% | 66.7 | 66.7 |
| Electron beam flatness constancy | 3% | 66.7 | 66.7 |
| X-ray and electron symmetry | 3% | 66.7 | 66.7 |
| Emergency off switches | Functional | 100 | 61.1 |
| Wedge, electron cone interlocks | Functional | 66.7 | 61.1 |
| Light / radiation field coincidence | 2 mm | 100 | 55.6 |
| Gantry / collimator angle indicators | 1° | 100 | 94.4 |
| Wedge position | 2 mm | 66.7 | 0 |
| Tray position | 2 mm | 100 | 0 |
| Applicator position | 2 mm | 66.7 | 0 |
| Field size indicators | 2 mm | 100 | 33.3 |
| Cross-hair centering | 2 mm diameter | 100 | 0 |
| Treatment couch position indicators | 2 mm / 1° | 66.7 | 11.1 |
| Latching of wedges, blocking tray | Functional | 66.7 | 16.7 |
| Jaw symmetry | 2 mm | 66.7 | 27.8 |
| Field light intensity | Functional | 66.7 | 0 |

Table 8. The comparison between the present state of QA procedures for linear accelerators presented by international organizations with that of HP_QAPs in Korea (annual QA)

| Procedure | Tolerance | Execution rate of the international organization(%) | Execution rate of the hospital in Korea (%) |
| --- | --- | --- | --- |
| X-ray / electron output calibration constancy | 2% | 100 | 50 |
| Field size dependence of X-ray output constancy | 2% | 66.7 | 50 |
| Output factor constancy for electron applicators | 2% | 100 | 50 |
| Central axis parameter constancy (PDD, TAR) | 2% | 100 | 100 |
| Off-axis factor constancy | 2% | 66.7 | 0 |
| Transmission factor constancy for all treatment accessories | 2% | 66.7 | 16.7 |
| Wedge transmission factor constancy | 2% | 66.7 | 50 |
| Monitor chamber linearity | 1% | 66.7 | 16.7 |
| X-ray output constancy vs gantry angle | 2% | 66.7 | 50 |
| Electron output constancy vs gantry angle | 2% | 66.7 | 50 |
| Off-axis factor constancy vs gantry angle | 2% | 66.7 | 0 |
| Arc mode | 1% from baseline | 66.7 | 0 |
| Safety interlocks | Functional | 66.7 | 16.7 |
| Collimator rotation isocenter | 2 mm diameter | 100 | 50 |
| Gantry rotation isocenter | 2 mm diameter | 100 | 50 |
| Table rotation isocenter | 2 mm diameter | 100 | 50 |
| Coincidence of collimator, gantry and table axis with the isocenter | 2 mm diameter | 100 | 50 |
| Coincidence of radiation and mechanical isocenter | 2 mm diameter | 66.7 | 50 |
| Table top sag | 2 mm | 66.7 | 33.3 |
| Vertical travel of table | 2 mm | 66.7 | 50 |

Table 9. The comparison between the present state of QA procedures for linear accelerators presented by western advanced countries and that of HP_QAPs in Korea (daily QA)

| Procedure | Tolerance | Execution rate of the developed countries (%) | Execution rate of the hospital in Korea (%) |
|---|---|---|---|
| X-ray output constancy | 3% | 100 | 0 |
| Electron output constancy | 3% | 100 | 0 |
| Localizing lasers | 2 mm | 100 | 64.7 |
| Optical distance indicator (ODI) | 2 mm | 75 | 64.7 |
| Door interlock | Functional | 75 | 94.1 |
| Audiovisual monitor | Functional | 100 | 100 |
| Beam status(ON) indicators | Functional | 75 | 94.1 |
| Room radiation monitors | Functional | 50 | 0 |
| Field size indicator | 2 mm | 50 | 11.8 |
| Dynamic wedge factors | 2% | 50 | 0 |

Table 10. The comparison between the present state of QA procedures for linear accelerators presented by western advanced countries and that of HP_QAPs in Korea (weekly QA)

| Procedure | Tolerance | Execution rate of the developed countries (%) | Execution rate of the Hospital in Korea (%) |
| --- | --- | --- | --- |
| X-ray output constancy | 2% | 75 | 83.8 |
| Electron output constancy | 2% | 75 | 83.3 |
| X-ray central axis dosimetry parameter constancy (PDD, TAR, TPR) | 2% | 50 | 0 |
| Electron central axis dosimetry parameter constancy (PDD) | 2 mm @ therapeutic depth | 50 | 0 |
| X-ray beam flatness constancy | 2% | 100 | 0 |
| Electron beam flatness constancy | 3% | 100 | 0 |

Table 11. The comparison between the present state of QA procedures for linear accelerators presented by western advanced countries and that of HP_QAPs in Korea (monthly QA)

| Procedure | Tolerance | Execution rate of the developed countries (%) | Execution rate of the Hospital in Korea (%) |
|---|---|---|---|
| X-ray output constancy | 2% | 50 | 83.3 |
| Electron output constancy | 2% | 50 | 83.3 |
| X-ray central axis dosimetry parameter constancy (PDD, TAR, TPR) | 2% | 100 | 33.3 |
| Electron central axis dosimetry parameter constancy (PDD) | 2 mm @ therapeutic depth | 100 | 33.3 |
| X-ray beam flatness constancy | 2% | 50 | 66.7 |
| Electron beam flatness constancy | 3% | 50 | 66.7 |
| X-ray and electron symmetry | 3% | 50 | 66.7 |
| Emergency off switches | Functional | 75 | 66.7 |
| Wedge, electron cone interlocks | Functional | 75 | 61.1 |
| Light / radiation field coincidence | 2 mm | 100 | 55.6 |
| Gantry / collimator angle indicators | 1° | 100 | 94.4 |
| Wedge position | 2 mm | 50 | 0 |
| Tray position | 2 mm | 50 | 0 |
| Applicator position | 2 mm | 50 | 0 |
| Field size indicators | 2 mm | 100 | 33.3 |
| Cross-hair centering | 2 mm diameter | 75 | 0 |
| Coincidence of collimator, gantry and table axis with the isocenter | 2 mm diameter | 50 | 94.4 |
| Treatment couch position indicators | 2 mm / 1° | 100 | 11.1 |
| Couch isocenter | 2 mm | 50 | 94.4 |
| Couch angle | 1° | 50 | 94.4 |

Table 12. The comparison between the present state of QA procedures for linear accelerators presented by western advanced countries and that of HP_QAPs in Korea (monthly QA)

| Procedure | Tolerance | Execution rate of the developed countries (%) | Execution rate of the Hospital in Korea (%) |
|---|---|---|---|
| X-ray / electron output calibration constancy | 2% | 100 | 50 |
| Field size dependence of X-ray output constancy | 2% | 75 | 50 |
| Output factor constancy for electron applicators | 2% | 75 | 50 |
| Central axis parameter constancy | 2% | 100 | 100 |
| Off-axis factor constancy | 2% | 50 | 0 |
| Transmission factor constancy for all treatment accessories | 2% | 75 | 16.7 |
| Wedge transmission factor constancy | 2% | 75 | 50 |
| Monitor chamber linearity | 1% | 50 | 16.7 |
| X-ray output constancy vs gantry angle | 2% | 75 | 50 |
| Electron output constancy vs gantry angle | 2% | 50 | 50 |
| Beam symmetry reproducibility vs gantry angle | 3% | 50 | 0 |
| Safety interlocks | Functional | 50 | 16.7 |
| Collimator rotation isocenter | 2 mm diameter | 100 | 50 |
| Gantry rotation isocenter | 2 mm diameter | 100 | 50 |
| Table rotation isocenter | 2 mm diameter | 100 | 50 |
| Coincidence of collimator, gantry and table axis with the isocenter | 2 mm diameter | 100 | 50 |
| Coincidence of radiation and mechanical isocenter | 2 mm diameter | 100 | 50 |
| Table top sag | 2 mm | 100 | 33.3 |